\documentstyle[prl,aps,multicol,epsf]{revtex}

\begin{document}

\title{Reply to Comment on: ``Rashba precession in quantum wire
with interaction''}
\author{Wolfgang H\"ausler
\thanks{On leave from: Physikalisches Institut,
Universit\"at Freiburg, D-79104 Freiburg, Germany}
}
\address{Fachbereich Physik, Universit\"at Siegen,
ENC, D-57068 Siegen \ and\\
I.\ Institut f\"ur Theoretische Physik der Universit\"at Hamburg,
Jungiusstr.\ 9, D-20355 Hamburg, Germany
}
\date{Received: \hspace*{3cm}}
\maketitle
\begin{abstract}
It is shown that the recent Comment \cite{yu} by Y. Yu on the
above article is not substantiated.
\end{abstract}

\begin{multicols}{2}
\narrowtext
Recently a Comment appeared on the ArXiv preprint server
\cite{yu} on my article `Rashba precession in quantum wires with
interaction' \cite{rashba-rapid}. Its main statement is to
question the dependence of the Rashba spin-orbit generated
persistent spin current on the interaction. I demonstrate now
that Comment \cite{yu} contains numerous faults, one of which
leading to the above (mis)conclusion.

The Author begins by stating that non-integer valued $J_\nu$
would contradict to the seminal work by Haldane \cite{haldane}.
In this regard \cite{rashba-rapid} carefully distinguishes
between eigenvalues and expectation values at non-zero $\alpha$
and it are the latter for which Eq.~(2) of \cite{yu} is correct
(as clearly stated before Eq.~(7) in \cite{rashba-rapid}). It
should be pointed out that the expression $\tilde J_\sigma$
found after Eq.~(9) in \cite{yu} is non-integer either
($q_R=m\alpha$ is tuned externally), contrary to what is stated
one sentence later.

Secondly, the Fermi velocities $v_{Fa}$ in \cite{yu} are taken
as spin dependent which in the effective mass description is
erroneous at given Fermi energy. The effective mass description
holds up to the perturbative order ${\cal O}(\alpha^5)$, as is
carefully derived in \cite{rashba-rapid}, cf.\ also
\cite{physe03}. The $v_{Fa}$ in \cite{yu} describe a
non-equilibrium situation.

The main error, however, occurs in Eq.~(10) of \cite{yu} where
the shifted current $\tilde J_\sigma$ is inserted into the
interaction, instead of the current $J_\sigma$. For not
explicitly spin dependent interactions the $J_\sigma=0$ state is
of course the state of lowest interaction energy and not
the state $\tilde J_\sigma=0$. Correct insertion would
immediately lead to the (correct) result, Eq.~(2) of \cite{yu}.
In Eq.~(10) of \cite{yu} the interaction depends on the Rashba
coupling $q_R$ while the Rashba spin-orbit term depends
explicitly on the interaction strength. Both is clearly not
representing the model under consideration, the Rashba
spin-orbit coupling being a single particle operator, cf.\
Eq.~(1) of \cite{rashba-rapid}.

In Eq.~(14) the prefactor $v_F/\lambda_{\sigma}$ has simply been
written {\em ad hoc} in front of the $J_{\sigma}$-linear term,
without further justification. Correct would have been this
term proportional to $q_R$ precisely as in Eq.~(14) with
$\lambda_{\sigma}=1$, describing indeed SU(2) symmetric spin sectors
(a case which has not been considered in \cite{wu}). It is not the
scope of this Reply to correct for the inconsistencies between
Eq.~(14) of \cite{yu} and the work by Y.-S.~Wu \cite{wu}, where
the Author of \cite{yu} is Coauthor.

In conclusion, the persistent spin current and the Rashba length
both being ground state properties (and as such unrelated to
$H_q$) {\em do} depend on the interaction. The Comment \cite{yu}
by Y.~Yu lacks substantiation.

\begin{raggedright}

\end{raggedright}
\end{multicols}
\end{document}